\documentclass{JINST}
\usepackage{graphicx,natbib,amssymb,lineno, epsf}

\title {High-Temperature Superconducting Level Meter for Liquid Argon Detectors}

\author{A.~Bueno, A.J.~Melgarejo\footnote{Now at Columbia University}, J.L.~Navarro, S.~Navas and A.G.~Ruiz \\
Dpto. de F{\'i}sica Te\'orica y del Cosmos \& C.A.F.P.E.,
Universidad de Granada, 18071 Granada, Spain\\}

\abstract
{
Capacitive devices are customarily used as probes to measure 
the level of noble liquids in detectors operated for neutrino studies and 
dark matter searches. In this work we describe the 
use of a high-temperature 
superconducting material as an alternative to control the level of 
a cryogenic noble liquid. Lab measurements indicate that the superconductor 
shows a linear behaviour, a high degree of stability and offers a very 
accurate determination of the liquid volume. This device is therefore 
a competitive instrument and shows several advantages over conventional 
level meters.
}


\begin{document}

% ==================================
\section{Introduction}
\label{sec:intro}
% ==================================

Research on fundamental physics can largely benefit from the use of
instruments based on noble liquids, since they show several appealing
features as detecting medium~\cite{Doke}. In fact, during the last years we have
witnessed an increase on the interest and use of devices having
liquefied noble gases as their sensitive material~\cite{Fabjan}. 
This innovative technique has a wide scope of application that 
spans from calorimeters for collider 
physics~\cite{D0,H1,LHC} to TPCs used for the study of neutrino
properties~\cite{t600} or direct dark matter
searches~\cite{Gaitskell,WARP,ArDM}. To guarantee optimal and well-controlled 
running conditions, these detectors must be complemented with auxiliary
instrumentation to gauge the temperature, pressure, purity and level
of the liquid~\cite{book1,book2}. Accurate measurements of the liquid
height can be accomplished using different sorts of level
meters~\cite{Sawada,Bruschi}. Following the work done in~\cite{Arkharov}, 
we propose the 
use of a high-temperature superconducting material (SC) as 
an alternative to measure the level of a detector filled with liquid
argon. The SC we have tested cannot be used in detectors that use xenon as
the sensitive material, given the higher boiling point of this noble
gas. For neon detectors a different sort of SC must be used to monitor
the level.

In the following sections we describe the
experimental setup, the performance of the level meter and the
advantages it shows with respect to devices previously discussed in
the literature.   

% ==================================
\section{Experimental setup}
\label{sec:setup}
% ==================================

In Figure~\ref{fig:SC-level-esquema} we show the different elements
of the experimental setup. They are:

\begin{description}
\item[$\bullet$ Cryostat]
We use a vacuum-insulated AGIL-2 from Air Liquide~\cite{AirL} as external vessel.
It is a cylindrical two-liter stainless-steel dewar, whose top is
closed by a cork lid. Once the 9 cm vessel is filled with liquid nitrogen,
the measured evaporation rate, mainly through the porous cork,
amounts to $\sim$10~mm/hour. This is small enough to perform long time 
data-taking runs. However the fact that the top lid does not guarantee an
optimal thermal insulation from the outside represents a clear
disadvantage. In Section~\ref{sec:measurements}, we show that 
as the liquid nitrogen evaporates, the gradient of temperatures
between the liquid surface and the top of the dewar is high enough 
to alter the behaviour of the SC. This spoils the accuracy of the
level measurement in case long ($\ge$ 3 cm) SC are used. This unwanted
effect will be sensibly reduced in case a cryostat with a better thermal 
insulation is used~\cite{us}. 
%------------------------------------------------------------
\begin{figure}
\leftmargin=2pc
\begin{center}
\includegraphics[width=8cm, clip]{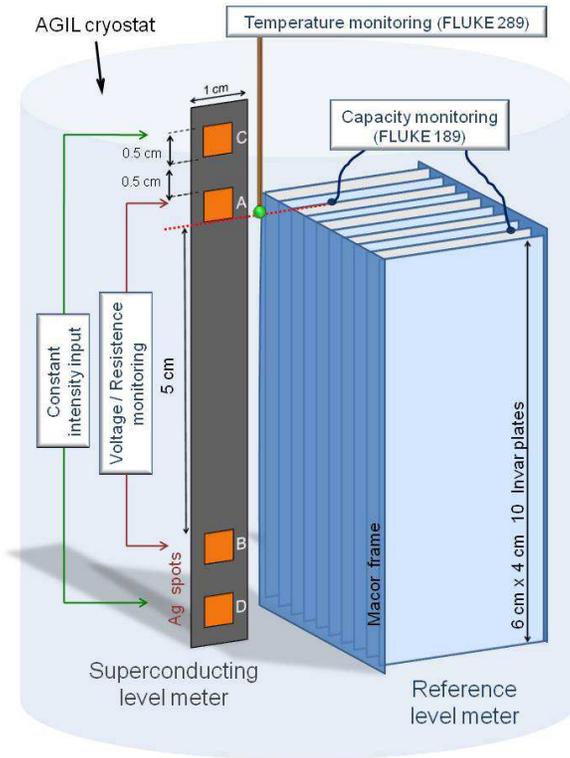}
\end{center}
\caption{Schematic drawing of the setup used for the superconductor (SC)
characterization. The reference capacity level meter,
the temperature probe and the superconductor are aligned together at the
level of the dotted line and immersed in a liquid nitrogen bath.
The SC data are obtained by measuring between points A and B
the resistance or the voltage. This last measurement 
requires the input of an external current.}
\label{fig:SC-level-esquema}
\end{figure}
%------------------------------------------------------------

\item[$\bullet$ The Coated Conductor]

The superconducting level meter is a flexible THEVA
DyBCO-tape~\cite{THEVA}. This coated conductor (CC) shows a layered
structure (see top panel of Figure~\ref{fig:sc-layers}). 
The main substrate is made of a thin (90~microns thick) 
high-strength Hastelloy C 276 steel tape. A double MgO layer follows. The first one
is an orientation layer deposited by inclined substrate deposition 
with a thickness of 3.7~microns; the second one is a high temperature
cap layer of 0.3~microns. A 1.6~micron superconducting DyBCO layer is finally deposited
by electron beam evaporation.

%------------------------------------------------------------
\begin{figure}
\leftmargin=2pc
\begin{center}
\includegraphics[width=6.5cm,angle=-90,clip]{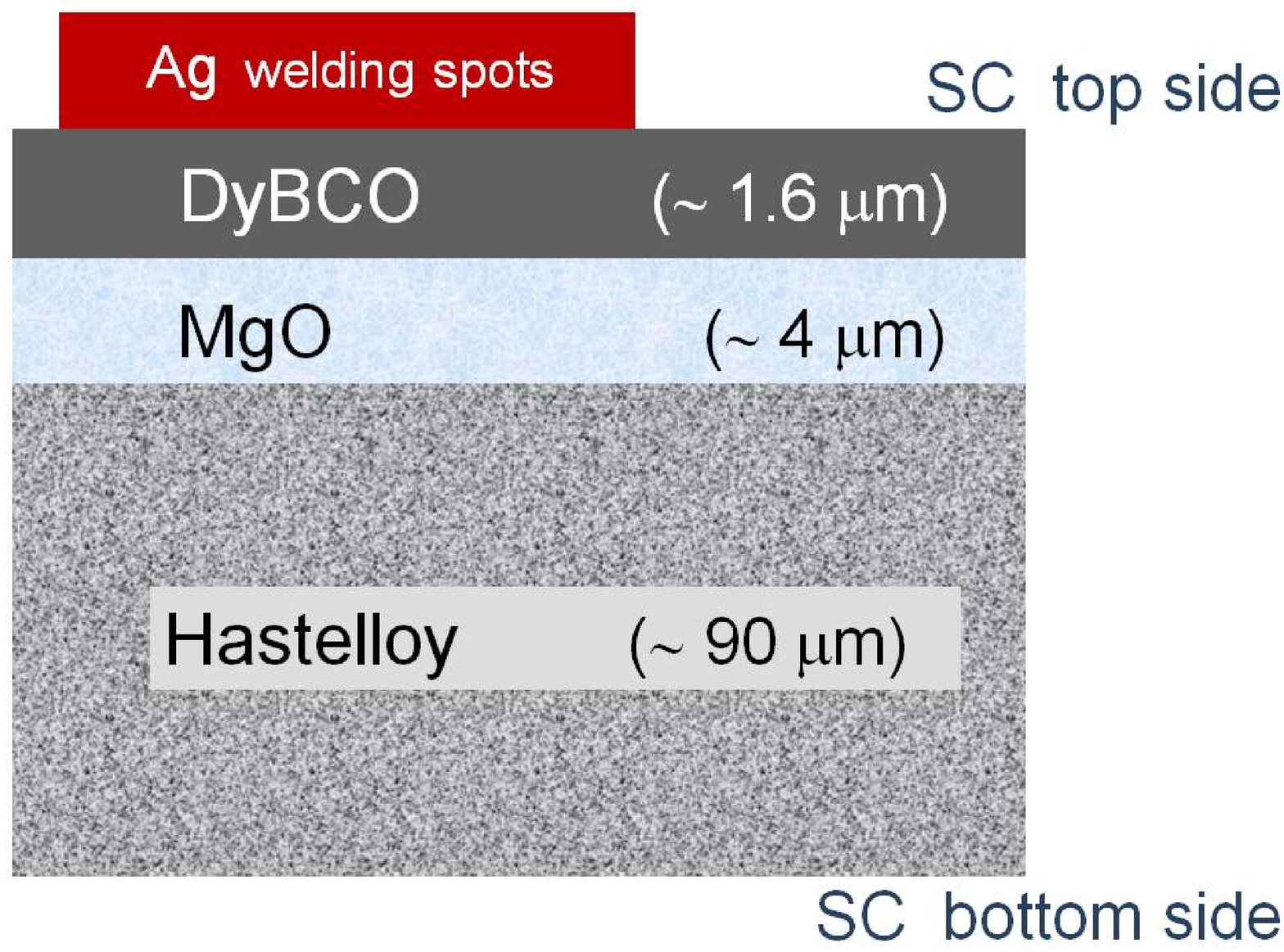} 
\includegraphics[width=8.5cm, clip]{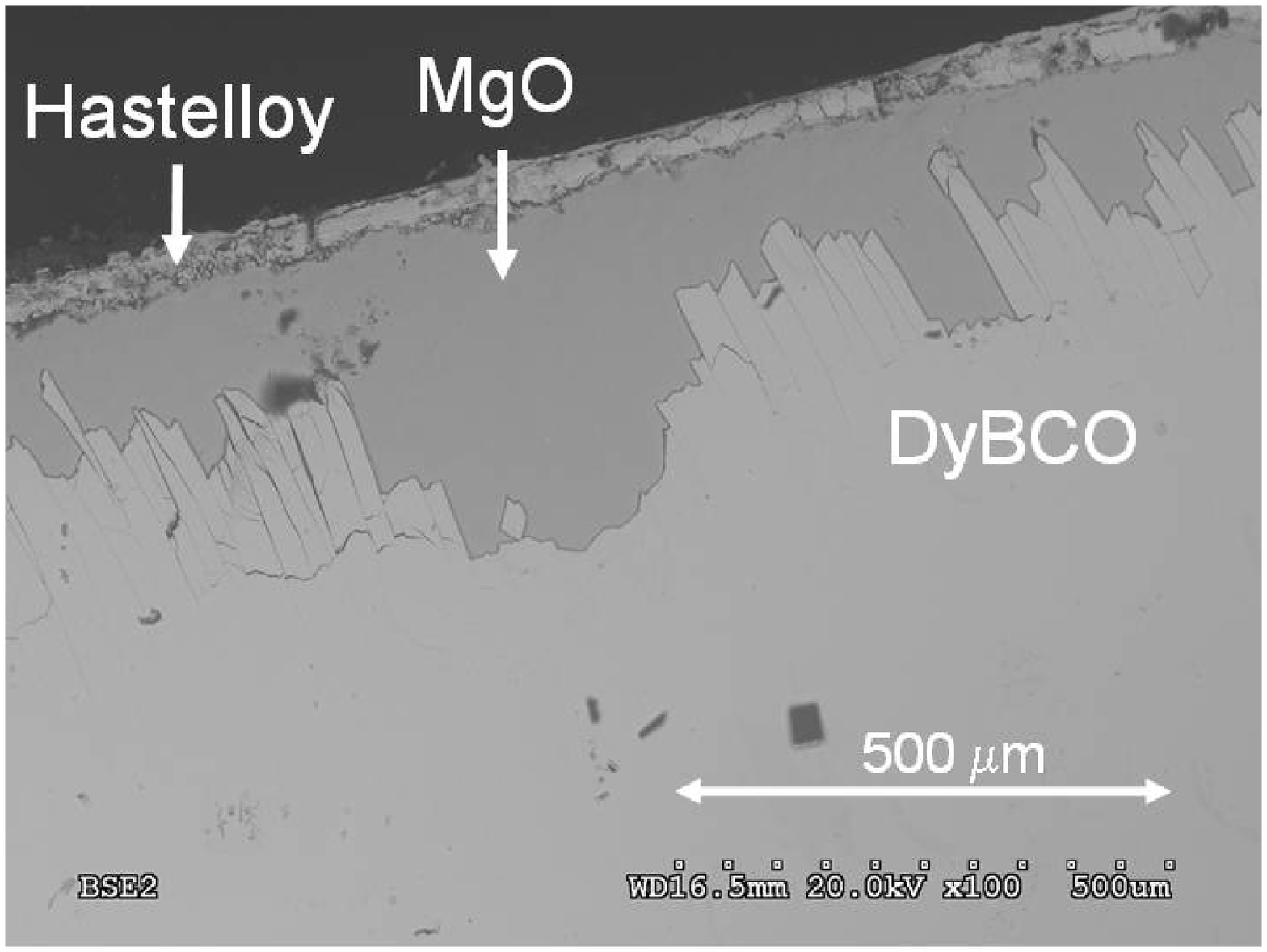} 
\end{center}
\caption{
Top:
Scheme of the coated conductor composition (Hastelloy tape substrate,
MgO buffered tape and superconducting DyBCO layer -SC-).
Some silver spots on top of it were added to allow
copper cable welding for signal readout.
Bottom:
Magnified picture of a coated conductor sample
taken with the Scanning Electron Microscope SEM/EDS technique.
The light grey region corresponds to the DyBCO layer
whereas the dark grey part below it is the MgO substrate.}
\label{fig:sc-layers}
\end{figure}
%------------------------------------------------------------

%------------------------------------------------------------
 \begin{figure}
 \leftmargin=2pc
 \begin{center}
 \includegraphics[angle=270,width=10cm,clip]{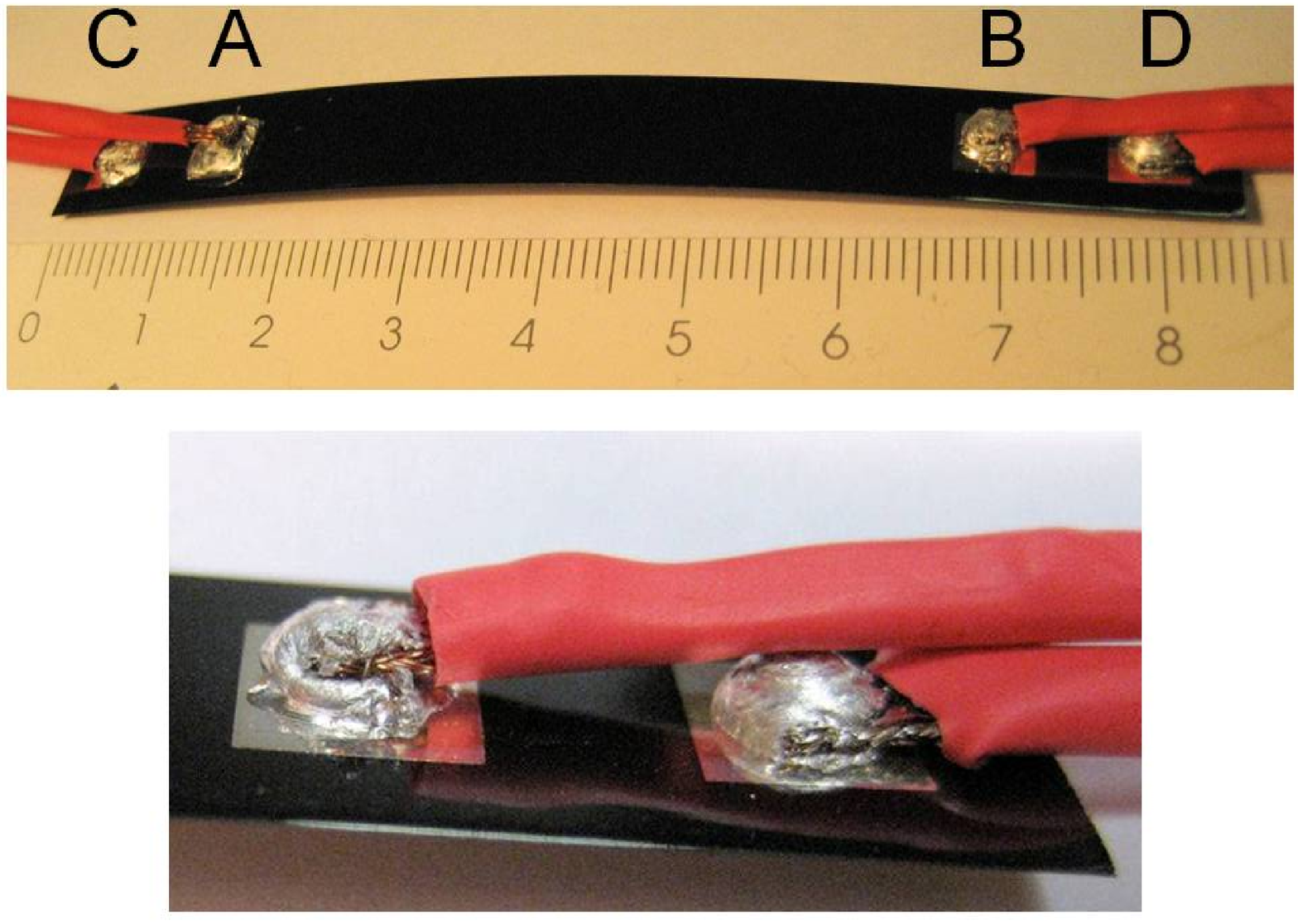} \\
 \end{center}
 \caption{Top: Picture of a finished DyBCO coated conductor ready for testing.
 The two external copper cables are connected to a stable current source
 ensuring a constant intensity flux. The two internal ones are
 used for voltage readout or a direct measurement of the resistance.
 Bottom: Detail of the copper--indium--silver soldered joints.}
 \label{fig:foto-SC}
 \end{figure}
%------------------------------------------------------------

We have tested six SC level meters. Each of them is a stripe
$\sim$8.5~cm long and 1~cm width (see top of
Figure~\ref{fig:foto-SC}). From the mechanical point of view, this 
device is simpler than the capacitive level meter used as reference 
(see below). We have analyzed the chemical composition
and spatial uniformity of the SC using a Scanning Electron Microscope (HITACHI S-3000N)
coupled to an Energy Dispersive X-Ray Spectrometer (BRUKER XFlash 3001). 
The idea was to look for irregularities, fractures or any other sort
of damage that could cause a potential failure of the SC. 
This scanning technique allows to examine and analyze samples at 
magnifications up to $\times$300,000.
The picture shown in the bottom panel of Figure~\ref{fig:sc-layers} 
was taken from the top of the level meter and close to one of the borders. 
It covers an area of about 1~mm$^2$. The three CC layers are clearly visible.
Figure~\ref{fig:Micro-histogram} shows an example of line scan analysis
done on the superconducting surface. A chemical study of
the substance reveals that, as expected, the Dy-Ba-Cu-O lines are
dominant. The high magnification surface topography taken along the surfaces
of the six manufactured SC level meters revealed a good uniformity in chemical
composition and the absence of irregularities or fractures in the
micro-structure.

To assess its performance, the SC level meter is carefully aligned to a multi-parallel-plate
capacitor (previously calibrated and used as reference level meter) and a temperature probe. 
Those elements are immersed in a liquid nitrogen (LN$_2$) bath ($\sim$
-195~$^{\circ}$C). The level of liquid nitrogen is inferred from the resistance value
R$_{SC}$ of the SC layer, which increases when
the LN$_2$ level decreases. This requires two electrical connections on the SC
(A and B in Figure~\ref{fig:foto-SC}) in case a multimeter is used to directly read
R$_{SC}$, and two more connections if it is obtained from the ratio of the measured voltage
when a constant current runs between pins C and D. In the latter case,
the SC is fed by a power supply (CPX200 from TT{\it i}~\cite{TTI}). 
To guarantee an optimal and stable electrical contact between the
measurement instruments and the superconducting layer, four 0.5$\times$0.5 cm$^2$ silver coated
squares were deposited at symmetric positions on the SC edges
(see bottom of Figure~\ref{fig:foto-SC}). Copper cables were 
finally soldered to the silver surface using indium (silver doped at 3$\%$).

All our tests have been performed using liquid nitrogen (boiling
point: 77 K). The SC still retains its superconducting nature at
temperatures as high as the argon boiling point (87 K). Therefore
similar results will be obtained in case we use liquid argon.
Moreover, the stripes appear to be robust and suitable for use in cryogenic detectors
since we observed that the SC structure was unaltered after being operated in subsequent
cycles of extreme temperature changes.

%------------------------------------------------------------
\begin{figure}
\leftmargin=2pc
\begin{center}
\includegraphics[width=9cm, clip]{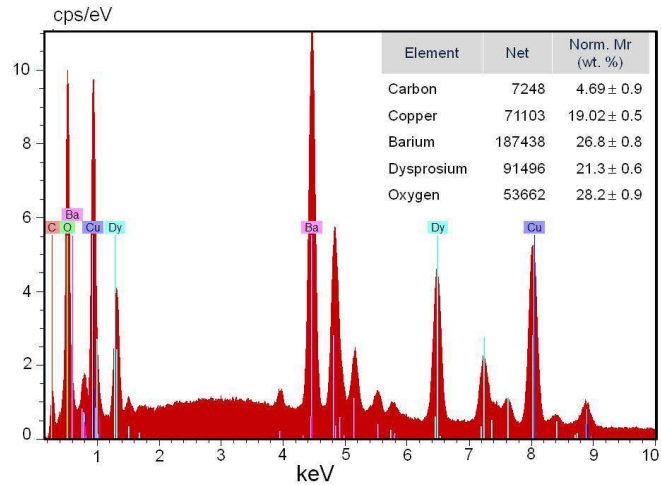}
\end{center}
\caption{Line scan on the CC DyBCO surface.The counting rate
per element and the corresponding relative abundance normalized
to the molecular weight are shown.}
\label{fig:Micro-histogram}
\end{figure}
%------------------------------------------------------------

\item[$\bullet$ Reference level meter]

The level of liquid nitrogen inside the dewar 
is continuously monitored with a capacitive level
meter~\cite{Laffranchi}. This custom-made device is 
a multi-parallel-plate capacitor composed of 10 Invar plates
40~mm (wide) $\times$  60~mm (high) and 1 mm thick 
(see Figure~\ref{fig:foto-level meter}).
The plates, spaced by 1.4 mm, are glued to a Macor supporting
structure. They work as anode and cathode alternatively.
The chosen structure aims at decreasing the errors in level measurement
by increasing the capacitance signal. 
Since the dielectric constant of the liquid ($\epsilon_{LN_2}$ = 1.434) 
and gas (can be considered as 1) are different, the capacity
changes linearly as a function of the area immersed inside the liquid
(see top Figure~\ref{fig:sc_level_linear}).

%------------------------------------------------------------
 \begin{figure}
 \leftmargin=2pc
 \begin{center}
 \includegraphics[width=6cm, clip]{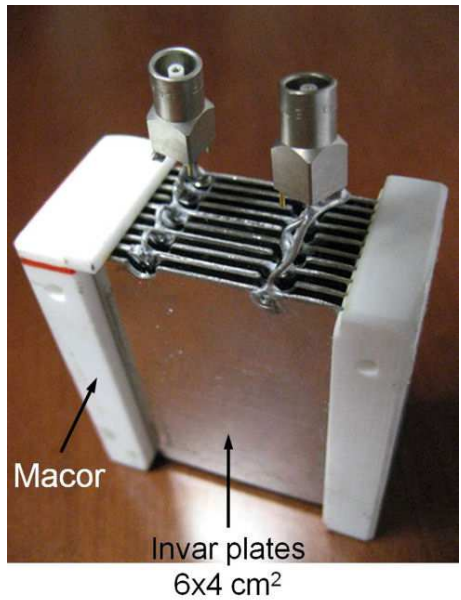}
 \end{center}
 \caption{Picture of the reference multi-parallel-plate capacity level meter.
The Invar plates (alternately connected) are supported by a Macor frame.}
 \label{fig:foto-level meter}
 \end{figure}
%------------------------------------------------------------

The level meter is connected to a FLUKE 189~\cite{FLUKE} multimeter to measure the
capacity in intervals of five seconds. The capacity variation from
initial conditions (level meter fully immersed in liquid nitrogen) 
to the final state (level meter in air) amounts to 44$\pm$1~pF. 
Combining the data from several runs, 
we obtain an average precision for the level measurement of 
350$\pm$50 microns~\cite{Carmen}.

\item[$\bullet$ Sensors and data readout]

Temperature variations inside the cryostat were monitored with a 
FLUKE 289 multimeter which uses a 80BK Integrated DMM Temperature Probe.
This Type-K probe allows to measure temperatures down to -200~$^{\circ}$C
with 0.1~$^{\circ}$C resolution. A second FLUKE 189 digital multimeter allows
to measure the SC resistance and the levelmeter capacity, as well. A KEITHLEY
electrometer (mod. 6514~\cite{Keithley}) measures the voltage between
the SC pins while the external power supply is on. All these data 
are stored in a hard drive. Off-line analyses will allow to obtain the
resistance of the SC in two different ways (see next Section). 
\end{description}

% ==================================
\section{Measurements and results}
\label{sec:measurements}
% ==================================

\begin{figure}
\leftmargin=2pc
\begin{center}
\includegraphics[width=8cm, clip]{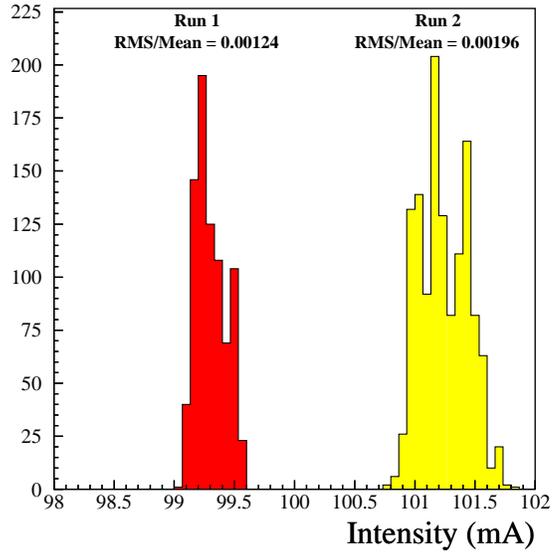}
\end{center}
\caption{
Example of input intensity fluctuations as provided by the external power supply.
The data, corresponding to two runs at I$_{CD}$ $\sim$ 100~mA, show intensity
fluctuations below 2~per mil.}
\label{fig:monitor-intensity}
\end{figure}

The working principle of the SC level meter is well known: 
While immersed in liquid nitrogen, the coated conductor retains its
superconducting nature. As soon as it goes above -183.0 $\pm$
0.5~$^{\circ}$C, the resistance increases causing the loss of 
the superconducting properties. Therefore a measurement of the SC
resistance can be translated into a measurement of the liquid level
inside the vessel. In this work, the SC resistance is measured 
using two different approaches:

\begin{itemize}

 \item {\it Digital Multimeter (DM) mode:} the digital multimeter is connected 
to the internal A--B pins and the resistance R$_{AB}$ read directly
(0.001~$\Omega$ resolution).

 \item {\it External Current (EC) mode:} an external power supply is connected to
pins C--D so that a constant current (I$_{CD}$)
runs through the SC layer. The voltage between pins A--B (V$_{AB}$) 
is measured with an electrometer and the effective resistance computed
(V$_{AB}$/I$_{CD}$). Measurements were taken for three different 
values of the current: 10, 50 and 100~mA. The measured I$_{CD}$ was
stable with fluctuations below the 2~per mil level (see Figure~\ref{fig:monitor-intensity}). 
For a current of 50 mA, we observed a maximum resistance of 4 $\Omega$ at the end 
of each data taking period.

\end{itemize}
%------------------------------------------------------------
%------------------------------------------------------------

%------------------------------------------------------------
\begin{figure}
\leftmargin=2pc
\begin{center}
\includegraphics[width=8cm, clip]{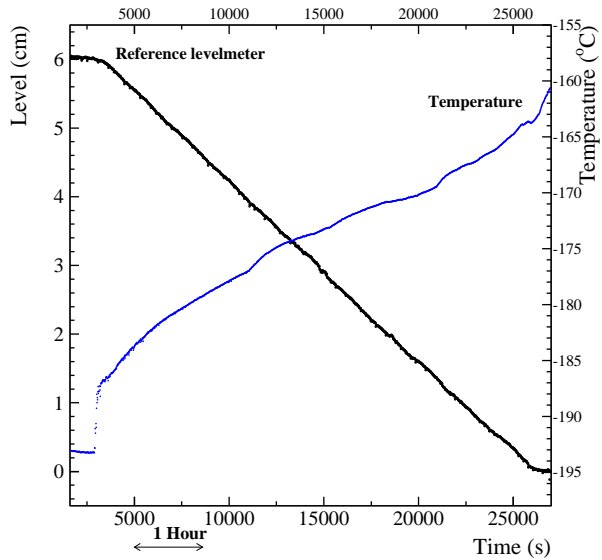}
\end{center}
\caption{Behaviour of the reference level meter (6--0~cm range) 
as a function of the temperature change.}
\label{fig:sc_level_temp}
\end{figure}
%------------------------------------------------------------

The calibration of the SC level meter (namely, the conversion 
of the measured SC resistance into units of length) is done by means
of a reference level meter. 
We believe this is a necessary step to be done prior to the installation of the SC in its final location.
These two elements, together with 
a temperature probe, are carefully aligned 
(dotted line in Figure~\ref{fig:SC-level-esquema}) and fixed to a
frame to avoid displacements during the measurement. 
The three devices are read continuously in time slots of 5 seconds. 
Typical runs extend for about 10~hours, the time needed to evaporate a
$\sim$9~cm column of liquid nitrogen.

The behaviour of the reference capacitive level meter is well
understood (see Figure~\ref{fig:sc_level_temp}): it shows 
two flat regions corresponding to the fully immersed condition
(leftmost part of the time axis) and dry state (rightmost part of the
abscissa). The central region corresponds to
the liquid nitrogen evaporation phase. The change in capacity is linear,
thus providing a direct, stable and precise indication of the liquid level.

The temperature probe also shows a smooth behaviour, increasing from
$\sim$-194~$^{\circ}$C to $\sim$-160~$^{\circ}$C. The increased rate
at high temperature (about 5.7~$^{\circ}$C/cm) and the observed fluctuations with time are
due to the poor thermal insulation offered
by the cork lid. 

\begin{figure}
\leftmargin=2pc
\begin{center}
\includegraphics[width=9cm, clip]{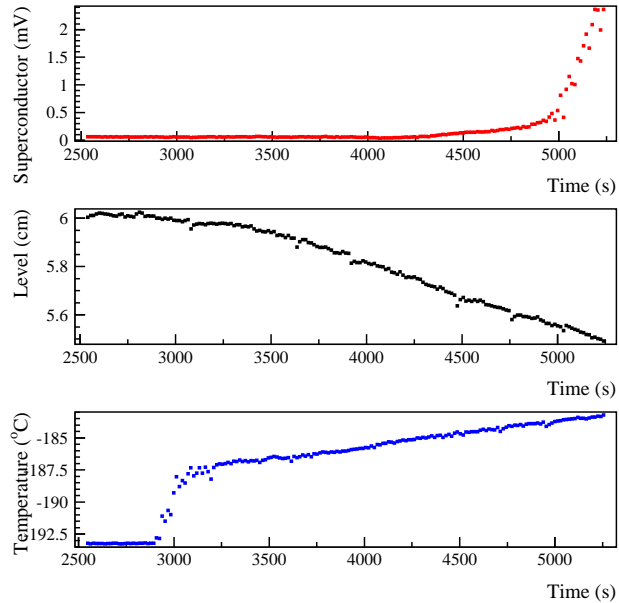}
\end{center}
\caption{Superconductor, reference level meter and temperature data
corresponding to a liquid level near the top region used for 
alignment purposes. The superconductor measurement was taken on 
{\it EC} mode at $I_{CD}$= 50 mA.}
\label{fig:sc_level_temp_zoom}
\end{figure}

The characterization of the six manufactured SC level meters was
carried out immersing each of them tens of times in liquid
nitrogen. Figure~\ref{fig:sc_level_temp_zoom} shows a zoom of the
top region of the experimental setup. The upper panel corresponds to
the SC level meter. Its resistance remains stable and very low ($\sim$0.18~$\Omega$)
while fully immersed in the liquid. It suddenly increases 
as the SC zone below pin A dries and warms up above the minimum
superconductivity temperature. The delay with respect to the
behaviour of the reference level meter has to do with the gradient of
temperatures above the surface of the liquid nitrogen. The SC
resistance only changes once its temperature is 
above -183~$^{\circ}$C. From the average of several measurements the 
offset among the two level meters amounts to 3.9$\pm$0.3~mm.
How the offset depends on the temperature gradient is being investigated
and will be reported elsewhere~\cite{us}.

\begin{figure}
\leftmargin=2pc
\begin{center}
\includegraphics[width=9cm, clip]{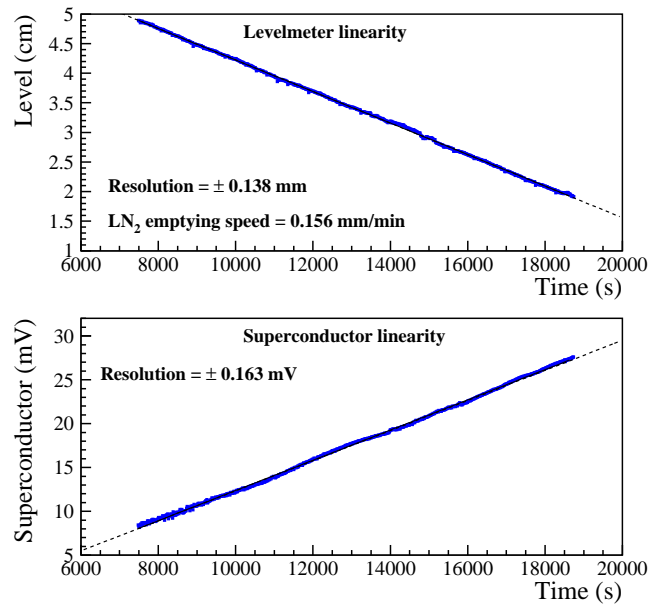}
\end{center}
\caption{Linearity of reference capacity level meter (top)
and superconductor (bottom) as function of time
(taken on {\it EC mode} at I$_{CD} = 50$~mA).
The solid lines correspond to straight line fits.}
\label{fig:sc_level_linear}
\end{figure}

\begin{figure}
\leftmargin=2pc
\begin{center}
\includegraphics[width=9cm, clip]{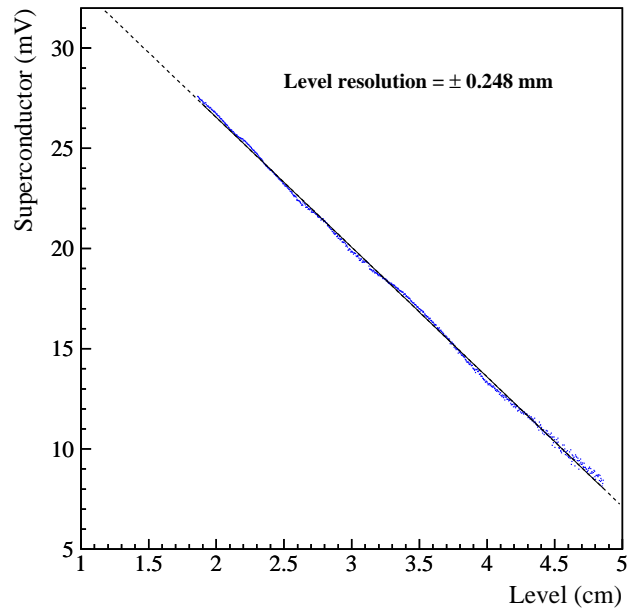}
\end{center}
\caption{Calibration curve: correlation between superconductor data
and reference capacity level meter (taken on {\it EC mode} at I$_{CD} = 50$~mA).
The resolution is computed as
the RMS deviation from the linear fit (solid line).}
\label{fig:sc_level_calibra}
\end{figure}

Figure~\ref{fig:sc_level_linear} shows how the resistance of the SC
level meter increases linearly for a long period of time ($\sim$4~h).
In our range of interest, the measured resistance grows up to 0.5 $\Omega$. 
The capacitive level meter shows a similar behaviour. The combination
of the data from the two level meters allows to obtain a calibration
for the SC (see Figure~\ref{fig:sc_level_calibra}). The root mean
square (RMS) from the line fit to the data gives the precision for the
level measurement. For the run depicted in the figure it amounts to
approximately 250 $\mu$m.

\begin{figure}
\leftmargin=2pc
\begin{center}
\includegraphics[width=9cm, clip]{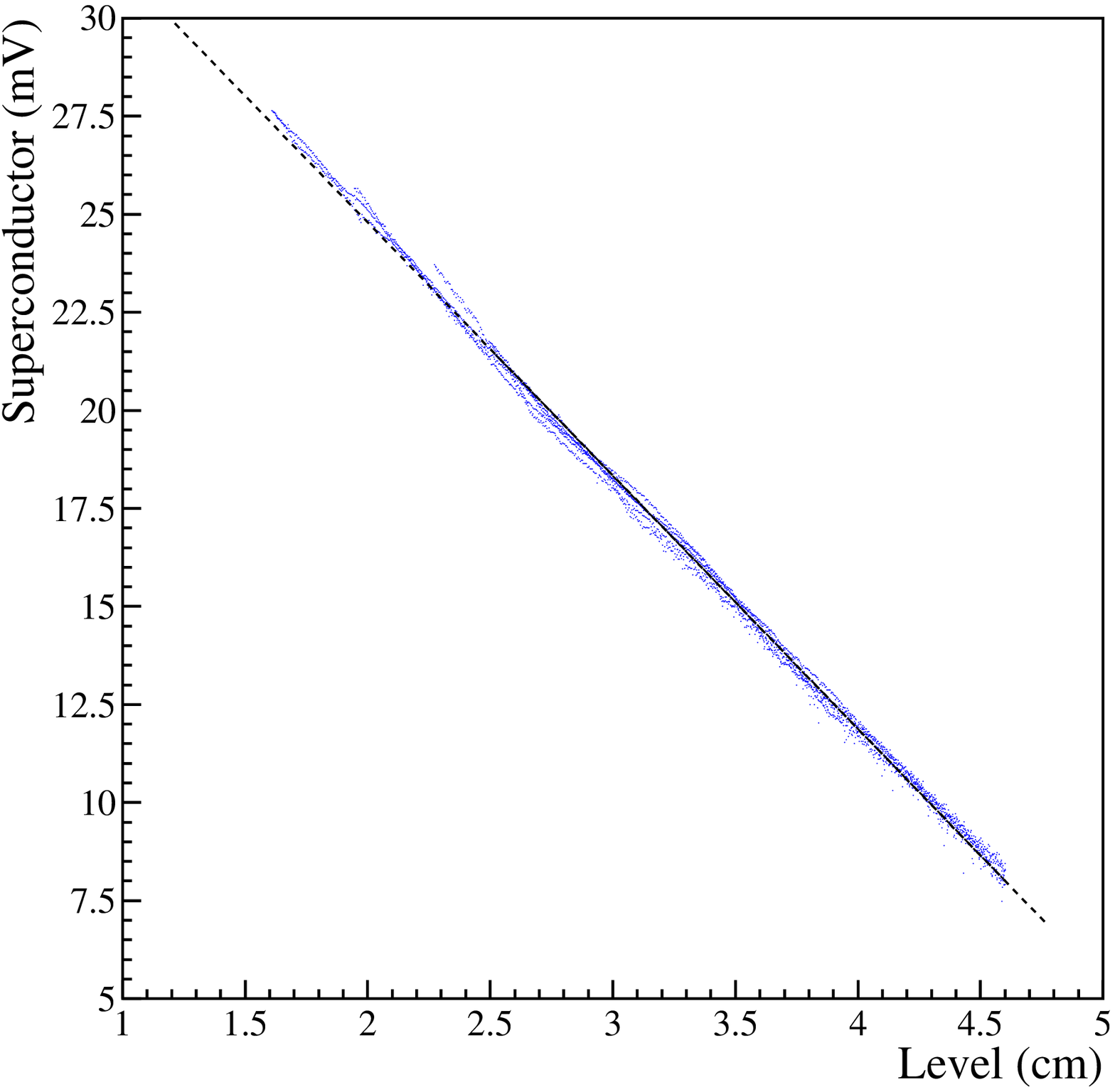}
\end{center}
\caption{Correlation between SC data and reference capacity level meter
for 8 runs on {\it EC mode} at I$_{CD} = 50$~mA.}
\label{fig:all_50ma}
\end{figure}

Run by run the reproducibility is good as can be seen from
Figure~\ref{fig:all_50ma}. However when the gradient of temperatures
inside the dewar is very high, the linear behaviour is lost. This
seriously limits the range of applicability of our measurements. The
poor thermal insulation forces us to restrict ourselves to just 
one third of the total length of the SC level meter. We hope that 
optimized experimental conditions will allow the use of longer SC
level meters. 

Despite the problems introduced by the high gradients of temperatures
we have found, our results are encouraging. Restricting ourselves to a
length of up to 2.5 cm, the outcomes are similar regardless of the 
sort of measurement performed (either {\it DM} or {\it EC} mode). For
the {\it EC} mode,
the computed resolutions do not exhibit a strong dependence 
with the intensity of the input current, as can be seen from
Table~\ref{tab:final}, where we show the expected resolutions for data
taken in {\it DM} mode and three different intensities (10, 50 and 100
mA) for the {\it EC} mode. The resolution is taken as the RMS from the
line fit done to all the runs corresponding to a particular mode of 
measurement. The expected resolutions go from 280 $\mu$m up to 
370 $\mu$m, and therefore are similar to the ones
obtained with the capacitive level meter.

\begin{table}
\begin{center}
\begin{tabular}{|lc|}\hline
Measurement & Resolution ($\mu$m)\\ \hline
{\it DM} mode & 300 \\
{\it EC} mode ($I_{CD}$= 10 mA) & 370 \\
{\it EC} mode ($I_{CD}$= 50 mA) & 280 \\
{\it EC} mode ($I_{CD}$= 100 mA) & 365 \\ \hline
\end{tabular}
\end{center}
\caption{Expected SC level meter resolutions for the different
measurement methods described in the text. }
 \label{tab:final}
 \end{table}

%------------------------------------------------------------
%------------------------------------------------------------

%------------------------------------------------------------
%------------------------------------------------------------

%------------------------------------------------------------
%------------------------------------------------------------

%------------------------------------------------------------
%------------------------------------------------------------

% ===============================
\section{Conclusions}
\label{sec:concl}
% ===============================

We have shown that superconducting materials are a promising
alternative to standard capacitive devices used to measure the level
of cryogenic liquids. With a simple
experimental setup, we have observed that the SC level meter is stable
and linear in the range of a few centimeters. The measurements show a
good reproducibility and the estimated precision is below
0.5 mm. We are confident that longer level meters will behave linearly
when tested under conditions with better thermal insulation. Compared 
to capacitive level meters, superconducting ones have several
advantages: they are cheaper, mechanically easier to mount and very 
simple electronics is needed to read them out.

% ===============================
\section{Acknowledgments}
\label{sec:acknowledgments}
% ===============================

This work has been carried out under the auspices of the Spanish Ministry of Education \& Science 
(Grant FPA2006-00684). The study of the SC composition has been carried out thanks to the 
invaluable help provided by Agust{\'\i}n Bueno (U. Alicante). We thank
A. Delgado (U. Granada) for useful discussions about superconductors.

% ===============================


\begin{thebibliography}{}
% ===============================
\bibitem{Doke}
T. Doke, 1981 Fundamental properties of liquid argon, krypton and xenon
as radiation media {\emph {Portugal Phys.} {\bf 12} 9.}

\bibitem{Fabjan}
M. Aleksa and C.W. Fabjan, Fundamental Physics with Noble Liquid
Detectors presented at the 15th IEEE International Conference on
Dielectric Liquids, Coimbra, Portugal; June 2005.

\bibitem{D0}
S. Abachi et al., 1994 The D0 detector
{\emph {Nucl. Instrum. Meth.} {\bf A338} 185.}

\bibitem{H1}
B. Andrieu et al., 1993 The H1 liquid argon calorimeter system
{\emph {Nucl. Instrum. Meth.} {\bf A336} 460.}

\bibitem{LHC}
ATLAS LAr Collaboration, 1996 LAr Technical Design Report CERN-LHCC-96-041.

\bibitem{t600}
S. Amerio et al., 2004 Design, construction and tests of the ICARUS T600 detector
{\emph {Nucl. Instrum. Meth.} {\bf A527} 329.}

\bibitem{Gaitskell}
 R J Gaitskell, 2004 Direct detection of dark matter 
{\emph {Ann. Rev. Nucl. Part. Sci.} {\bf 54} 315.}

\bibitem{WARP}
 R Brunetti et al., 2005 WARP liquid Argon detector for dark matter survey 
{\emph {New Astron. Review} {\bf 49} 265.}

\bibitem{ArDM}
 M Messina et al., A Status report of ArDM project presented at the
9th ICATPP Conference on Astroparticle, Particle, Space Physcis, Detectors
and Medical Physics Applications, Como, Italy; Oct 2005.

\bibitem{book1}
G.K. White and P. Meeson, 2002 Experimental Techniques in
Low-Temperature Physics, 4th Edition, Oxford University Press, New
York.
\bibitem{book2}
F. Pobell, 2007 Matter and Methods at Low Temperatures, 3rd Edition,
Springer-Verlag, Berlin.

\bibitem{Sawada}
R. Sawada et al., 2003 Capacitive level meter for liquid rare gases
{\emph {Cryogenics} {\bf 43} 449.}

\bibitem{Bruschi}
L. Bruschi, G. Delfitto and G. Mistura, 1999 Level meter for dielectric liquids
{\emph {Rev. Sci. Instrum.} {\bf 70} 1514.}

\bibitem{Arkharov}
I.A. Arkharov and V.Yu. Emel'yanov, 2000 High-Temperature superconducting 
level gauge for cryogenic liquid {\emph {Chem. Petrol. Eng.} {\bf 36} 542.}

\bibitem{AirL}
  AIR LIQUIDE,
\emph{http://www.airliquide.com}
\href{http://www.airliquide.com}
{\emph{}}

\bibitem{us}
The construction of a 50 liter cryostat, whose metal
top lid is sealed to reduce to a minimum the heat transfer, is under
way. It will be used to check, in optimal experimental conditions, 
the behaviour of long SC when immersed in liquid Argon. Results will be reported elsewhere.

\bibitem{THEVA}
  THEVA,
\emph{http://www.theva.com}
\href{http://www.theva.com}
{\emph{}}

\bibitem{TTI}
  Thurlby Thandar Instruments,
\emph{http://www.tti-test.com}
\href{http://www.tti-test.com}
{\emph{}}

\bibitem{Laffranchi}
M. Laffranchi, 2005 Test of a Liquid Argon Time Projection Chamber in a Magnetic Field
\href{http://neutrino.ethz.ch/diplomathesis.html}
{\emph {PhD. Thesis Diss. ETH No. 16002}.}

\bibitem{FLUKE}
  FLUKE,
\emph{http://www.fluke.com}
\href{http://www.fluke.com}
{\emph{}}

\bibitem{Carmen}
M.C. Carmona, 2008
{\emph {PhD. Thesis, University of Granada}.}

\bibitem{Keithley}
  KEITHLEY,
\emph{http://www.keithley.com}
\href{http://www.keithley.com}
{\emph{}}

\end{thebibliography}
\end{document}